\documentclass[aps,a4paper,showpacs,superscriptaddress,twocolumn]{revtex4}
\usepackage{tensor}
\usepackage{graphicx}
\usepackage{amsmath}
\usepackage{amssymb}
\usepackage{enumerate}
\usepackage{subfigure}
\usepackage{tabularx}
\usepackage[colorlinks=true, pdfstartview=FitV, linkcolor=blue, citecolor=red, urlcolor=black, breaklinks=true]{hyperref}
\newcommand{\be}{\begin{equation}}
\newcommand{\ee}{\end{equation}}
\newcommand{\ben}{\begin{eqnarray}}
\newcommand{\een}{\end{eqnarray}}
\newcommand{\bes}{\begin{subequations}}
\newcommand{\ees}{\end{subequations}}
\def\bal#1\eal{\begin{align}#1\end{align}}
\newcommand{\vphi}{\varphi}

\newcommand{\LL}{{\mathcal L}}

\begin{document}
\title{Maxwell-Higgs vortices with internal structure}
\author{D. Bazeia}\affiliation{Departamento de F\'\i sica, Universidade Federal da Para\'\i ba, 58051-970 Jo\~ao Pessoa, PB, Brazil}
\author{M.A. Marques}\affiliation{Departamento de F\'\i sica, Universidade Federal da Para\'\i ba, 58051-970 Jo\~ao Pessoa, PB, Brazil}
\author{R. Menezes}
\affiliation{Departamento de Ci\^encias Exatas, Universidade Federal
da Para\'{\i}ba, 58297-000 Rio Tinto, PB, Brazil}
\affiliation{Departamento de F\'\i sica, Universidade Federal da Para\'\i ba, 58051-970 Jo\~ao Pessoa, PB, Brazil}
\begin{abstract}
\end{abstract}
\date{\today}

\begin{abstract}
Vortices are considered in relativistic Maxwell-Higgs systems in interaction with a neutral scalar field. The gauge field interacts with the neutral field via the presence of generalized permeability, and the charged and neutral scalar fields interact in a way dictated by the presence of first order differential equations that solve the equations of motion. The neutral field may be seen as the source field of the vortex, and we study some possibilities, which modify the standard Maxwell-Higgs solution and include internal structure to the vortex.
\end{abstract}
\maketitle

{\it Introduction.} Vortices are planar structures that attain interesting topological behavior and have a diversity of applications in high energy physics and in condensed matter. In high energy physics in particular, in the case of a relativistic field theory, the Maxwell-Higgs model is perhaps the standard model that supports vortex configurations, as firstly shown by Nielsen and Olesen \cite{NO} and then by other researches \cite{B,VS,ViS}.

The standard Maxwell-Higgs model describes an Abelian gauge field $A_\mu$ minimally coupled to a charged scalar field $\vphi$ and obeys the local $U(1)$ symmetry. To develop vortex solutions, the model has to be enlarged to accommodate a potential of the Higgs type that develops spontaneous symmetry breaking. This model was long ago enlarged to accommodate the $U(1)\times U(1)$ symmetry, now with two gauge fields and two complex scalar fields that interact via an extension of the Higgs-like potential \cite{W}. An interesting result of this model was the possibility of adding internal structure to the solution, having superconducting properties. In \cite{VV} and in the more recent works \cite{super1,super2,super3} and in references therein one finds other studies related to the presence of superconducting strings.

Another line of investigation which also deals with the $U(1)\times U(1)$ symmetry concerns the study of a visible $U(1)$ gauge field sector $A_\mu$  and another hidden $U(1)$ gauge field sector $C_\mu$ that interact via the two gauge field tensors $F_{\mu\nu}=\partial_\mu A_\nu-\partial_\nu A_\mu$ and $G_{\mu\nu}=\partial_\mu C_\nu-\partial_\nu C_\mu$. The presence of the hidden sector is motivated by supersymmetric extensions of the standard model and by superstring phenomenology and may somehow play a role in the study of dark matter. Studies on vortex in such models appeared before in \cite{Sha1,Sha2}, and in references therein.

Recently, in \cite{ana1} we started a program to describe vortex structures in generalized models in $(2,1)$ spacetime dimensions, and in \cite{ana2} we studied the case of analytic vortex solutions. Other investigations on vortices that enlarge the $U(1)$ symmetry to accommodate new fields appeared before in \cite{r1,r2,S1,S2}, and more recently in \cite{r3,r4} and in references therein. In particular, in \cite{S1,S2} the $U(1)$ symmetry is enlarged to become $U(1)\times SO(3)$, to accommodate the $SO(3)$ spin group that under specific circumstances may lead to vortex solutions that behave as spin vortices. In this case, the $SO(3)$ symmetry is driven by the addition of neutral scalar fields that couple to the $U(1)$ symmetry via the charged Higgs-like field. 

These works motivated us to go further and investigate extended versions of the generalized model. Our ultimate goal is to deal with the case in which the $U(1)\times U(1)$ symmetry plays the basic role. In the current work, however, we follow another route and take the symmetry $U(1)\times Z_2$, coupling $U(1)$ to $Z_2$ symmetry via the addition of a neutral scalar field, with the coupling modulated by the presence of generalized permeability. The inclusion of the $Z_2$ symmetry which is controlled by the neutral field is perhaps the simplest possibility to enlarge the $U(1)$ symmetry, and below we show that it may modify the profile of the vortex in a way of current interest.

{\it The model.} We work in $(2,1)$ flat spacetime dimensions with the Lagrangian density
\be\label{lmodel}
	\LL = - \frac{1}{4}P(\chi)F_{\mu\nu}F^{\mu\nu} +|D_\mu\vphi|^2 + \frac12\partial_\mu\chi\partial^\mu\chi- V(\chi,|\vphi|)
\ee
where $\chi$ is a real scalar field, the neutral field, $\vphi$ is a complex scalar field, the charged field, and $A_\mu$ is the Abelian gauge field. Also, $F_{\mu\nu}=\partial_{\mu}A_\nu-\partial_{\nu}A_\mu$ is the electromagnetic tensor and $D_\mu=\partial_{\mu}+ie A_{\mu}$ stands for the covariant derivative. The potential is denoted by $V(\chi,|\vphi|)$ and may present terms that mix the real and complex scalar fields. We suppose $P(\chi)$ is a nonnegative function of the real scalar field and use the metric tensor $\eta_{\mu\nu}=(1,-1,-1)$ and $\hbar=c=1$. The equations of motion associated to the Lagrangian density \eqref{lmodel} are
\bes\label{geom}
\begin{align}
\partial_\mu\partial^\mu\chi + \frac14 P_\chi F_{\mu\nu}F^{\mu\nu} + V_\chi &=0\\
 D_\mu D^\mu \vphi + \frac{\vphi}{2|\vphi|}V_{|\vphi|} &=0,\\ \label{meqsc}
 \partial_\mu \left(P F^{\mu\nu}\right) &= J^\nu,
\end{align}
\ees
where the current is $J_{\mu} = ie(\overline{\vphi} D_{\mu} \vphi-\vphi\overline{D_{\mu}\vphi})$ and $P_\chi=dP/d\chi,$ $V_\chi = \partial V/\partial\chi$, and $V_{|\vphi|}=\partial V/\partial|\vphi|$. By setting $\nu=0$ in equation \eqref{meqsc}, one can show that for static field configurations the Gauss' law is satisfied with $A_0=0$. In this case, the vortex is electrically neutral since its electric charge vanishes. 

To search for topological solutions, we consider static configurations and suppose that
\be\label{ansatz}
\chi=\chi(r),\quad \vphi = g(r)e^{in\theta}, \quad \vec{A} = -{\frac{\hat{\theta}}{er}(a(r)-n)},
\ee
in which $n\in\mathbb{Z}$ is the vorticity. The functions $\chi(r)$, $a(r)$ and $g(r)$ obey the boundary conditions
\bal\label{bc1}
\chi(0) &=\chi_0, & g(0) &=0, & a(0)&=n,\\
\chi(\infty) &=\chi_\infty, & g(\infty)&=v, & a(\infty)&=0.\label{bc2}
\eal
Here, $\chi_0$, $\chi_\infty$ and $v$ are parameters involved in the symmetry breaking of the potential. Considering the fields described by equations \eqref{ansatz}, the magnetic field has to satisfy
\be\label{b}
B = -F^{12} = -\frac{a^\prime}{er},
\ee
where the prime stands for the derivative with respect to $r$. By using this, one can show the magnetic flux is quantized
\be
\Phi =2\pi\int rdr B = \frac{2\pi n}{e}.
\ee
The equations of motion \eqref{geom} with the static fields \eqref{ansatz} assume the form
\bes\label{secansatz}
\begin{align}
\frac{1}{r} \left(r \chi^\prime\right)^\prime &= P_\chi\frac{{a^\prime}^2}{2e^2r^2} + V_\chi, \\ 
\frac{1}{r} \left(r g^\prime\right)^\prime &= \frac{a^2g}{r^2} + \frac12 V_{|\vphi|}, \\
r\left(P\frac{a^\prime}{er}\right)^\prime &= 2eag^2.
\end{align}
\ees

The energy density for static field configurations can be calculated standardly; one uses \eqref{ansatz} to write
\be\label{rhoans}
\rho= P \frac{{a^\prime}^2}{2e^2r^2} +  {g^\prime}^2 + \frac{a^2g^2}{r^2} + \frac12{\chi^\prime}^2 + V.
\ee
The equations of motion \eqref{secansatz} are of second order and present couplings between the fields. In order to get first order equations, we use the Bogomol'nyi procedure \cite{B} and introduce an auxiliary function $W=W(\chi)$ to write the energy density \eqref{rhoans} as
\be
\begin{aligned}
	\rho &= \frac{P(\chi)}{2}\left(\frac{a^\prime}{er} \pm \frac{e(v^2-g^2)}{P(\chi)}\right)^2 +  \left(g^\prime \mp \frac{ag}{r} \right)^2 \\
	     &\hspace{4mm} + \frac12\left(\chi^\prime \mp \frac{W_\chi}{r}\right)^2 + V -\left(\frac{e^2}{2}\frac{\left(v^2-g^2\right)^2}{P(\chi)} + \frac12\frac{W^2_\chi}{r^2} \right) \\
	     &\hspace{4mm} \pm \frac{1}{r} \left( W - a\left(v^2-g^2\right)\right)^\prime,
\end{aligned}
\ee
where $W_\chi=dW/d\chi$. If the potential is written as
\be\label{potgen}
V(\chi,|\vphi|) =  \frac{e^2}{2}\frac{\left(v^2-|\vphi|^2\right)^2}{P(\chi)} + \frac12\frac{W^2_\chi}{r^2},
\ee
the energy becomes
\be
\begin{aligned}\label{ene}
	E &= 2\pi\int_0^\infty r\,dr\,\frac{P(\chi)}{2}\left(\frac{a^\prime}{er}\pm  \frac{e(v^2-g^2)}{P(\chi)}\right)^2\\
	&\hspace{4mm}+ 2\pi\int_0^\infty r\,dr\,\left(g^\prime \mp \frac{ag}{r} \right)^2 \\
	  & \hspace{4mm} + 2\pi\int_0^\infty r\,dr\, \frac12\left(\chi^\prime \mp \frac{W_\chi}{r}\right)^2 + E_B,
\end{aligned}
\ee
where 
\be\label{ebogo}
\begin{split}
	E_B &= \pm 2\pi\int_0^\infty dr\,\left( W - a\left(v^2-g^2\right) \right)^\prime \\
	    &= 2\pi\left|W(\chi(\infty))-W(\chi(0))\right| + 2\pi|n|v^2.
\end{split}
\ee
Since the three integrands in the energy \eqref{ene} are all non-negative, we see that the energy is bounded by $E_B$, i.e., $E\geq E_B$. If the solutions obey the equations
\be  
\chi^\prime = \pm\frac{W_\chi}{r}\label{fophi}
\ee
and
\bes\label{foe}
\bal\label{fog}
g^\prime &= \pm\frac{ag}{r}, \\ \label{foa}
-\frac{a^\prime}{er} &=\pm \frac{e\left(v^2-g^2\right)}{P(\chi)},
\eal
\ees
the Bogomol'nyi bound is saturated, such that the energy is minimized to $E=E_B$. Therefore, we have obtained three first order equations to study the problem, since they satisfy the equations of motion \eqref{secansatz}. As one knows, the fact that the solutions of the above first order equations \eqref{fophi} and \eqref{foe} saturate the Bogomol'nyi bound implies stability against decay into similar lower energy configurations.

It is worth commenting that the equation for the real scalar field \eqref{fophi} does not depend on the other fields. Thus, the real scalar field can be seen as a source to generate the vortex configuration, and we call it the source field. Although this is not apparent from the equations of motion \eqref{secansatz}, it is clear in the first order equations. Moreover, concerning the first order equations, it seems that the model one is dealing with is the bosonic portion of a larger, supersymmetric theory, which will be further investigated elsewhere. Here we keep working with the above model, since it unviels several interesting possibilities of investigations of current interest. 

An interesting issue concerns the presence of the radial coordinate in the first order equation \eqref{fophi}, which follows from the Bogomol'nyi procedure to minimize the energy of the static field configurations. One notes that the potential $V(\chi,|\vphi|)$ gained an extra contribution, the last term in the right hand side of equation \eqref{potgen}, to close the Bogomol'nyi procedure. This extra contribution remind us very much of the modification introduced before in \cite{prl} to circumvent the Derrick-Hobard scaling theorem \cite{H,D}, which inform us that a real scalar field, described by standard Lagrangian, cannot support stable static solution unless we work with a single spatial dimension.

To avoid instability of static solutions in $(2,1)$ spacetime dimensions, in \cite{prl} we suggested taking the real scalar field model in the form
\be\label{rmodel}
{\cal L}=\frac12\partial_\mu\chi\partial^\mu\chi-\frac12\frac{W_\chi^2}{r^2},
\ee
with the potential changed exactly as it has appeared above in the last term in equation \eqref{potgen}. We then see that if one considers
$g=|\vphi|\to v$ and $a\to0$, one gets to the minimum energy configuration for the gauge and complex scalar fields, and the model \eqref{lmodel} changes to the real scalar field model \eqref{rmodel}. In this case, the model \eqref{rmodel} is governed by the first order equations \eqref{fophi},
as shown before in \cite{prl}. Moreover, in the case one considers $\chi\to\tilde\chi$, with $\tilde\chi$ being a constant, a minimum of the scalar field potential that obeys $W_\chi(\tilde\chi)=0$, and since $P(\tilde\chi)$ is a positive real constant, one sees that the model \eqref{lmodel} becomes the standard Maxwell-Higgs model and obeys the first order equations \eqref{foe}, with $\chi\to\tilde\chi$, so
it has the same solutions, after rescaling the radial coordinate as 
\be 
r\to r\, \sqrt{P(\tilde\chi)\,}.  
\ee
Below we consider $P(\chi)$ such that $P(\tilde\chi)=1$, so we will not need to rescale the radial coordinate.

Let us now work with the energy density \eqref{rhoans}, which can be written as magnetic, gradient and potential contributions
\be
\rho = \rho_{mag} + \rho_{grad_\vphi} + \rho_{grad_\chi} + \rho_{pot},
\ee
where
\bes\label{rhop}
\bal
 \rho_{mag} &= P(\chi) \frac{{a^\prime}^2}{2e^2r^2} \\
 \rho_{grad_\vphi} &= {g^\prime}^2 + \frac{a^2g^2}{r^2} \\
 \rho_{grad_\chi} &=\frac12{\chi^\prime}^2\\
 \rho_{pot} &= \frac{e^2}{2}\frac{\left(v^2-g^2\right)^2}{P(\chi)} + \frac12\frac{W^2_\chi}{r^2}.
\eal
\ees
The first order equations \eqref{fophi} and \eqref{foe} allow us to write $\rho_{pot} = \rho_{mag} + \rho_{grad_\chi}$ and $ \rho_{grad_\vphi} = 2{g^\prime}^2$, so we can write 
\be
\rho = 2\rho_{mag} + \rho_{grad_\vphi} + 2\rho_{grad_\chi}.
\ee
This equation is interesting because it separates the energy density of the vortex from the one of the source field, which are respectively given by 
\be\label{rhos}
\rho_{vortex}=2\rho_{mag} + \rho_{grad_\vphi} \quad\text{and}\quad \rho_{scalar}=2\rho_{grad_\chi}.
\ee
It is worth mentioning that, according to result \eqref{ebogo}, the energy of the source field is fixed for a given $W(\chi)$. This implies that the same occurs for the vortex, in a manner that the magnetic and gradient portions may change, but its energy is always $2\pi|n|v^2$.

From now on, we work with dimensionless fields, keeping in mind that the rescale
\be
\begin{aligned}
	\vphi &\to v \vphi, & \chi &\to v\chi, & A_\mu &\to v A_\mu, \\
	 r &\to r/ev, & W_\chi &\to v W_\chi, & \LL &\to e^2v^4 \LL
\end{aligned}
\ee 
can be done. We further set $e=1$ and $v=1$, and work with unit vorticity, $n=1$, for simplicity.

\begin{figure}[htb!]
\centering
\includegraphics[width=4.2cm]{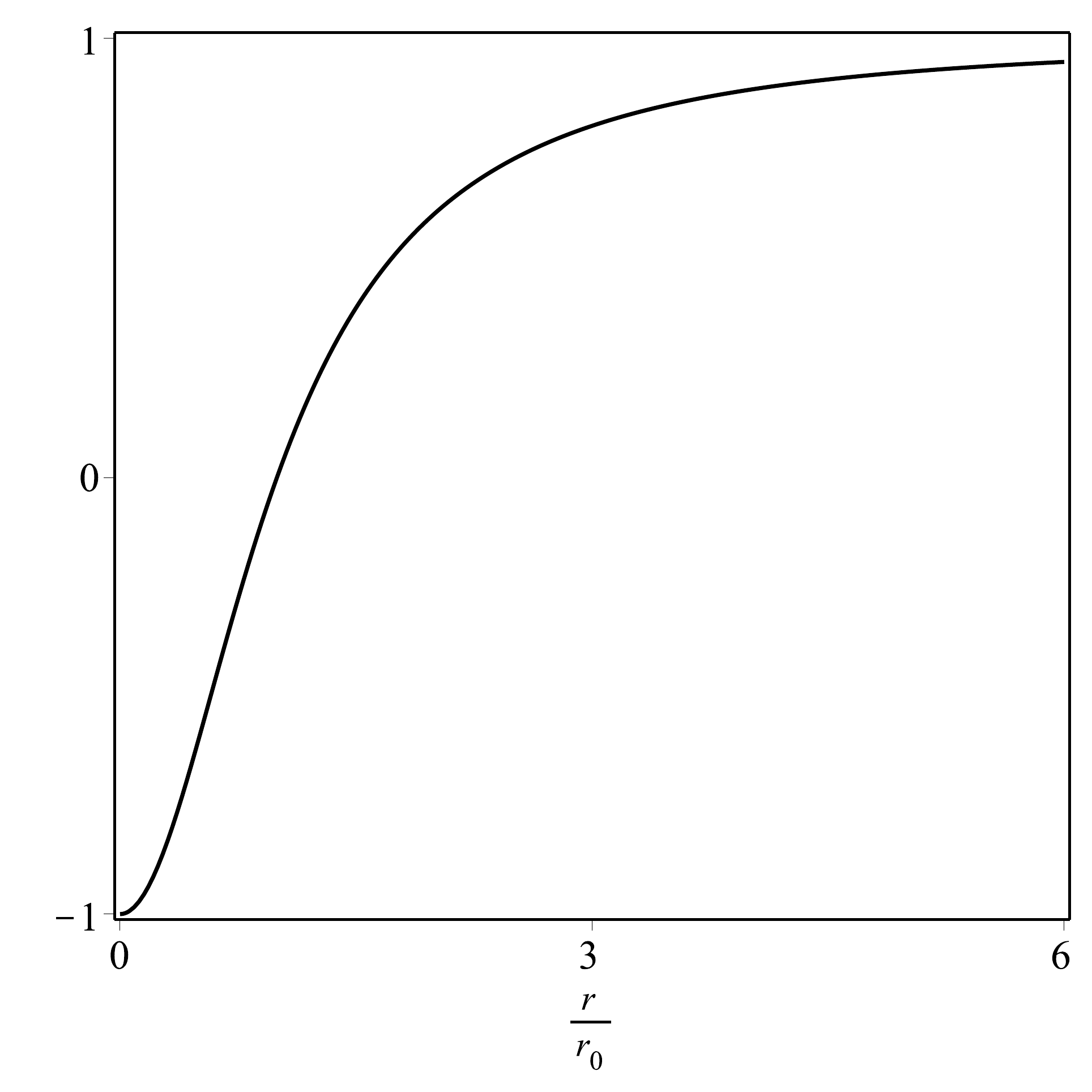}
\includegraphics[width=4.2cm]{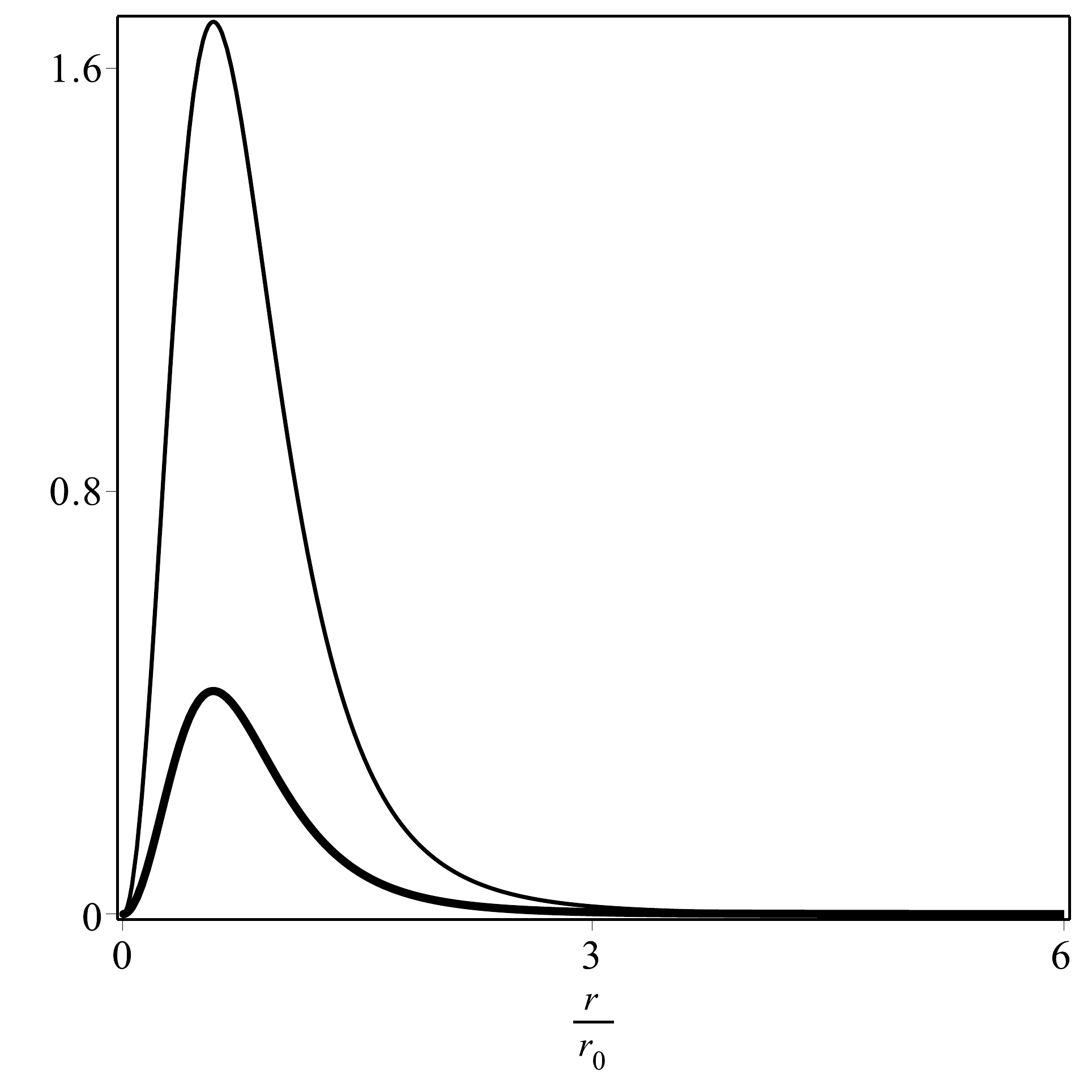}
\caption{The real scalar field solution \eqref{solphi} with the positive sign (left) and its energy density \eqref{rhosource} in terms of $r/r_0$ for $r_0=1$ and $2$, with the thickness of the lines increasing with $r_0$.}
\label{fig1}
\end{figure} 

{\it The source field.} Since the source field is independent from the other fields, we firstly deal with it and consider
\be\label{w}
W(\chi) = \chi-\frac{\chi^3}{3},
\ee
which was investigated in Ref.~\cite{prl} and more recently in \cite{BDR} and in \cite{BRR} to model planar skyrmion-like configurations, and also in \cite{BM} to study how massless Dirac fermions may behave in the background of such neutral planar structures. We can also use other possibilities, but here we consider the above $W(\chi)$, which changes the first order equation \eqref{fophi} to the form
\be
\chi^\prime = \pm\frac{1}{r} \left(1-\chi^2\right).
\ee
In this case, $\tilde\chi$ which we commented on below equation \eqref{rmodel} can be $\pm1$, and identify the fields $\chi_0$ and $\chi_\infty$ which appears in the boundary conditions \eqref{bc1} and \eqref{bc2}. The analytical solutions are given by
\be\label{solphi}
\chi(r) = \pm\frac{r^2-r_0^2}{r^2+r_0^2},
\ee
where $r_0$ is an arbitrary positive constant such that $\chi(r_0)=0$. The corresponding energy density is given by
\be\label{rhosource}
\rho_{scalar} = \frac{16\,r_0^4\, r^2}{(r_0^2 + r^2)^4}
\ee
In Fig.~\ref{fig1}, we depict the solution \eqref{solphi} (with the positive sign) and the energy density \eqref{rhosource} in terms of $r/r_0$. We see that the solution $\chi(r)$ varies smoothly from $-1$ at the origin to $1$ as $r$ increases to larger and larger values. This behavior is important to model the function $P(\chi)$ which controls the magnetic permeability of the model. The peaks in the right panel in Fig.~\ref{fig1} are at
$\bar{r}=r_0/\sqrt{3}$. Also, we observe that the energy density varies less significantly for higher values of $r_0$. This behavior will also appear for the vortex configurations that we study below. By integrating the above equation \eqref{rhosource}, we get the energy
\be  
E_{scalar} = \frac{8\pi}{3},
\ee
as expected from equation \eqref{ebogo}. Since the energy of the vortex equals $2\pi$, we conclude the total energy of the system is $E=14\pi/3$.

{\it The vortex profile.} We now go further and use the solutions \eqref{solphi} to model the magnetic permeability of the vortex. We first consider the possibility
\be
P_1(\chi)=\frac{1}{1-\chi^2},
\ee
which engenders the $Z_2$ symmetry. As we see, at $r= r_0$, the scalar field vanishes and $P_1(\chi)$ becomes unity, leading us with a vortex solution of the Nielsen-Olesen type \cite{NO}. However, the above choice makes $P_1(\chi)$ divergent at the origin and asymptotically. As we will show below, the singular behavior at the origin will be compensated by the vanishing of $B(r)$ as $r$ approaches the origin.  

\begin{figure}[t]
\centering
\includegraphics[width=4.2cm]{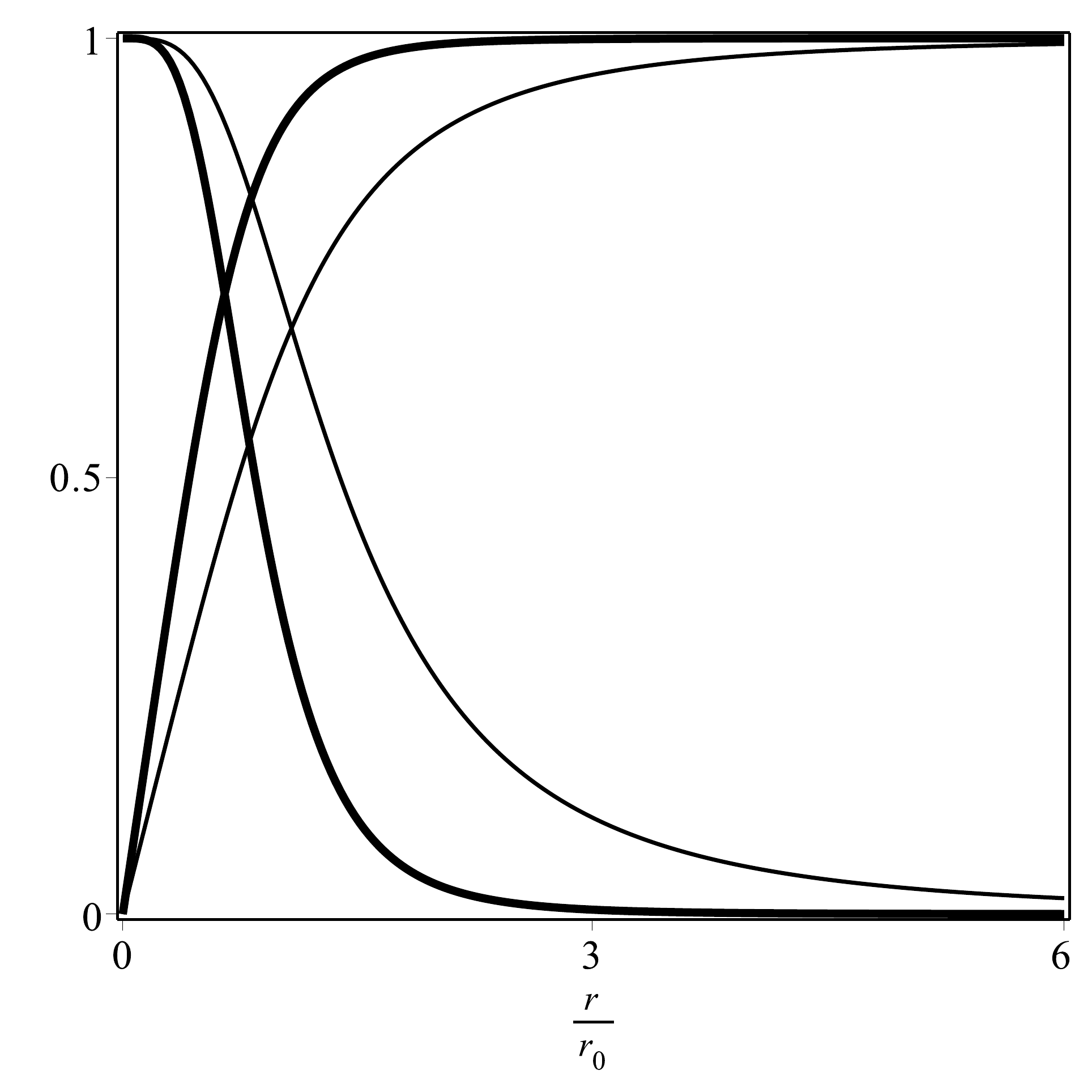}
\includegraphics[width=4.2cm]{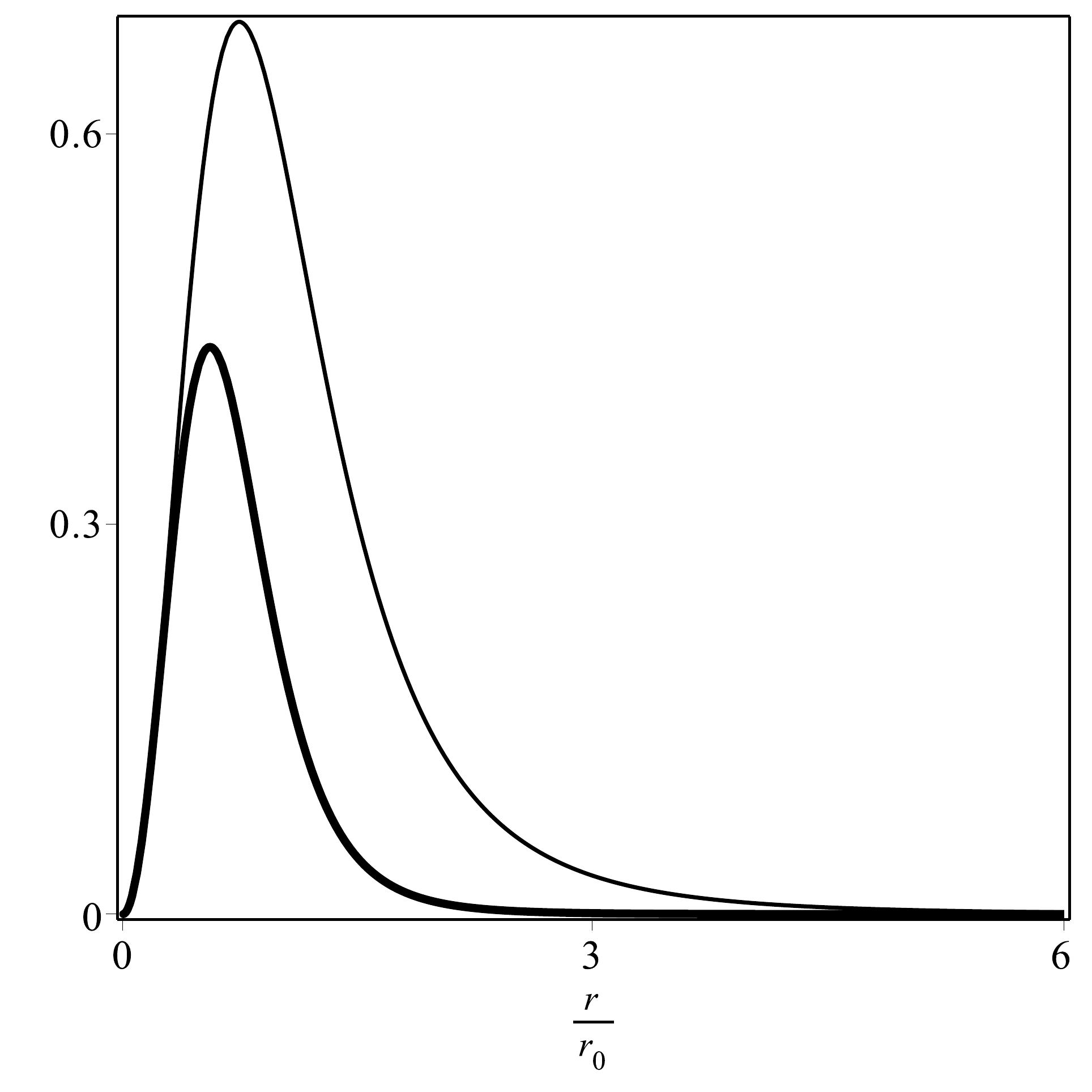}
\caption{The vortex solutions (left panel) $a(r)$ (descending lines) and $g(r)$ (ascending lines) and the magnetic field (right panel) in terms of $r/r_0$ for $r_0 =1$ and $2$, with the thickness of the lines increasing with $r_0$.}
\label{fig2}
\end{figure} 

To solve for the vortex, we consider the first order equations \eqref{fophi} and \eqref{foe} with the upper signs, and the solution \eqref{solphi} with the positive sign. The source field modifies the first order equation \eqref{foa} to become  
\be
-\frac{a^\prime}{r} = \frac{2\,r_0^2\left(1-g^2\right)}{r_0^2+ r^2},
\ee
which must be solved together with the equation \eqref{fog}. The solutions $g(r)$ and $a(r)$ are parametrized by the constant $r_0$ and can be evaluated numerically. They are depicted in Fig.~\ref{fig2} and are similar to the standard Nielsen-Olesen vortex configurations, but with the gauge field configuration $a(r)$ having a plateau near its core. This modifies the magnetic field in a significant way, which is also shown in the right panel in Fig.~\ref{fig2}. The magnetic field vanishes at the origin, showing a behavior which is important to avoid the divergence of $P(\chi)$ and control the energy of the solution. This is different from the standard vortex solution \cite{NO} and remind us very much of the behavior of the magnetic field in the Chern-Simons-Higgs model \cite{JW}; see also \cite{Ko,Ba}, which investigates the Maxwell-Higgs model, modified to incorporate generalized magnetic permeability, but with no extra neutral field.

\begin{figure}[t!]
\centering
\includegraphics[width=4.2cm]{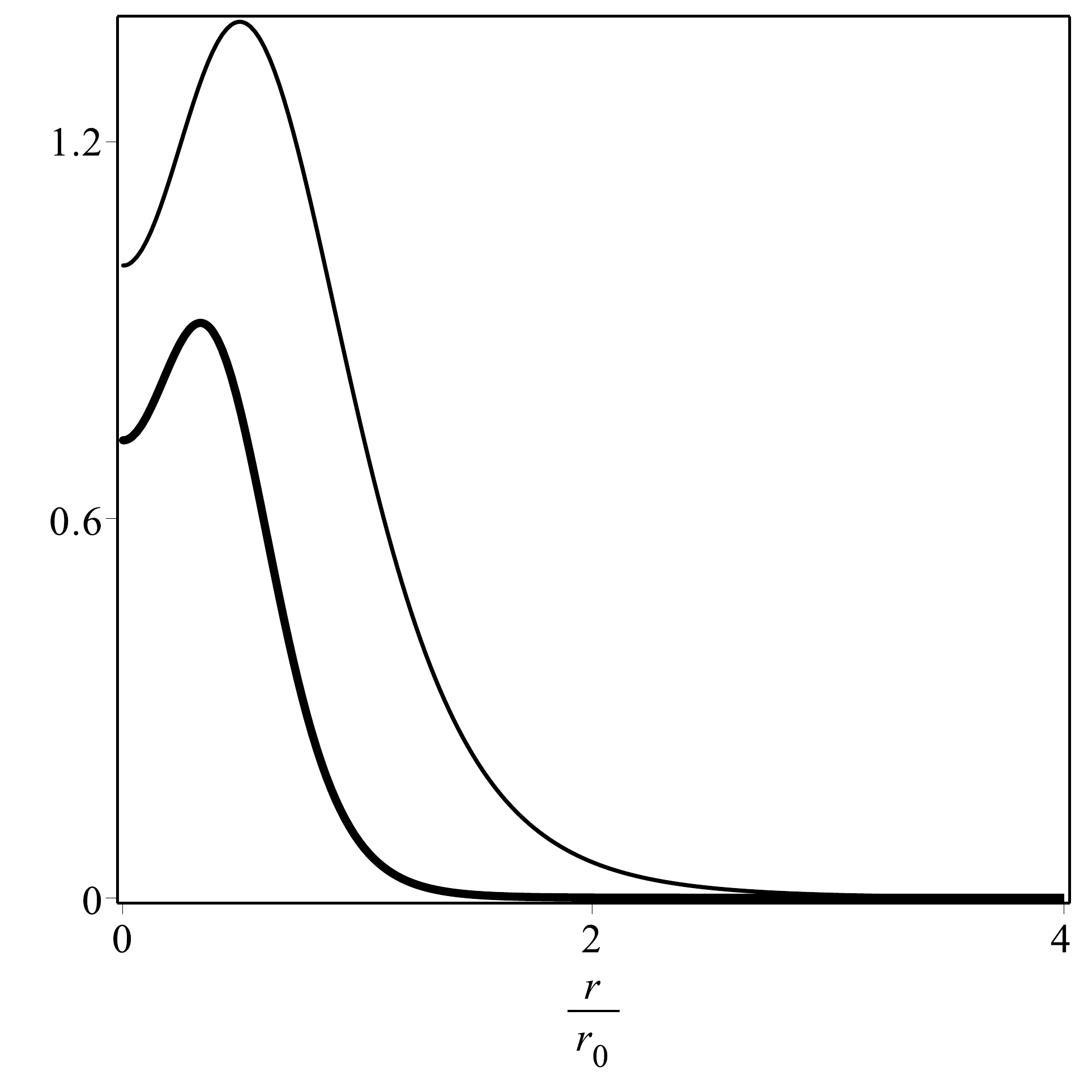}
\includegraphics[width=4.2cm]{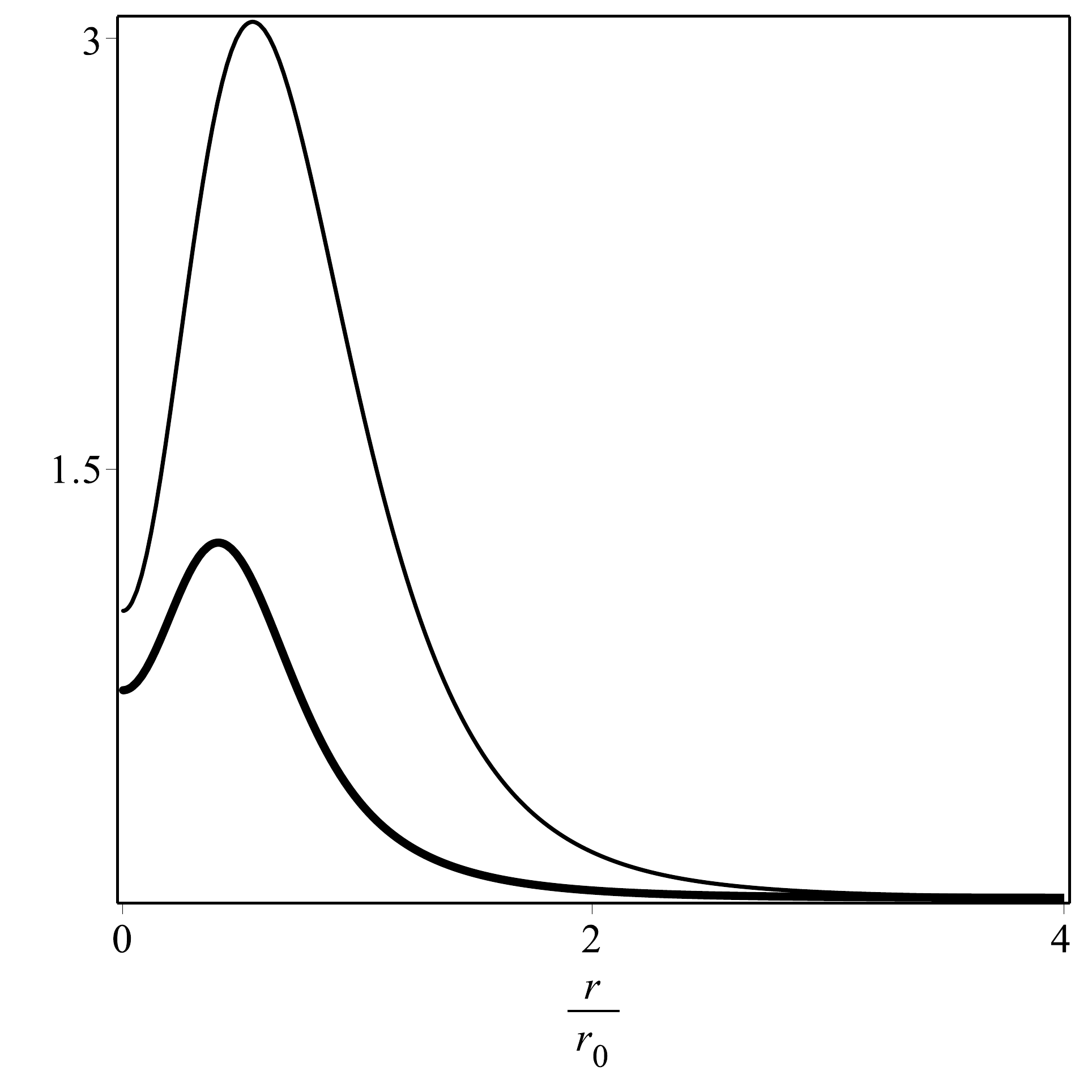}
\caption{The energy densities of the vortex (left panel) and the total energy density (right panel) for $r_0 = 1$ and $2$, with the thickness of the lines increases with $r_0$.}
\label{fig3}
\end{figure}

In Fig.~\ref{fig3}, we depict the energy density of the vortex and the total energy density of the field configurations for $r_0= 1$ and $2$.
We see that as $r_0$ increases, both the position and the height of the maximum of the energy densities decrease. 

\begin{figure}[htb!]
\centering
\includegraphics[width=4.2cm]{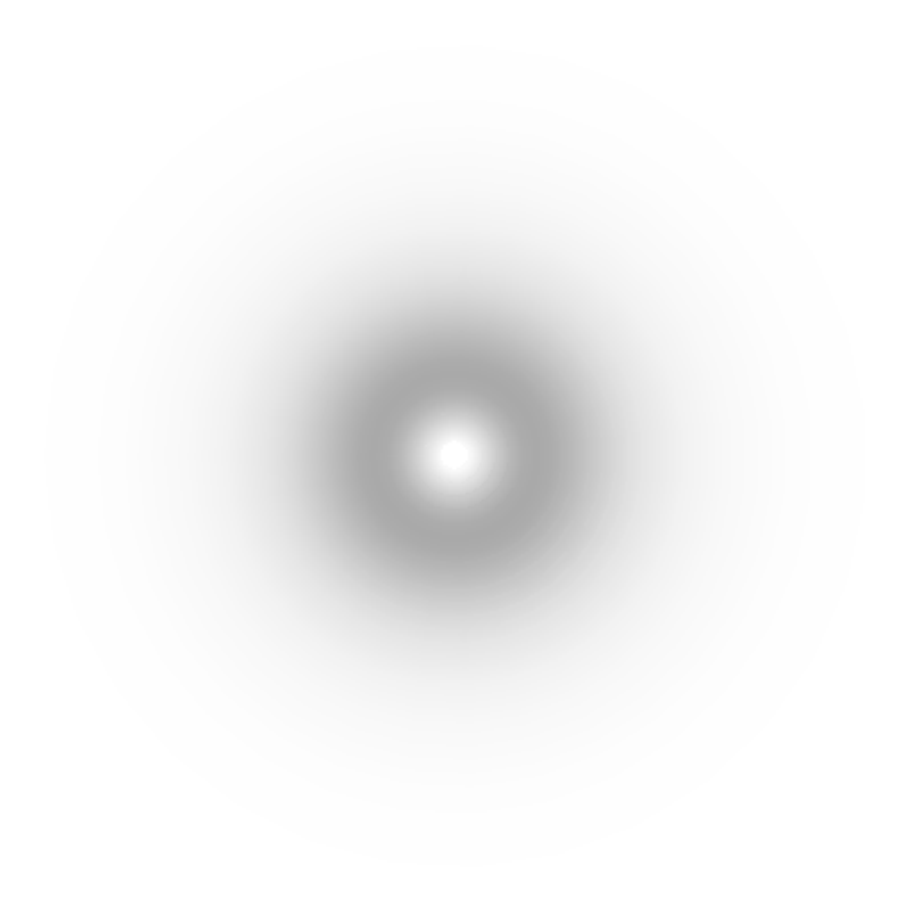}
\includegraphics[width=4.2cm]{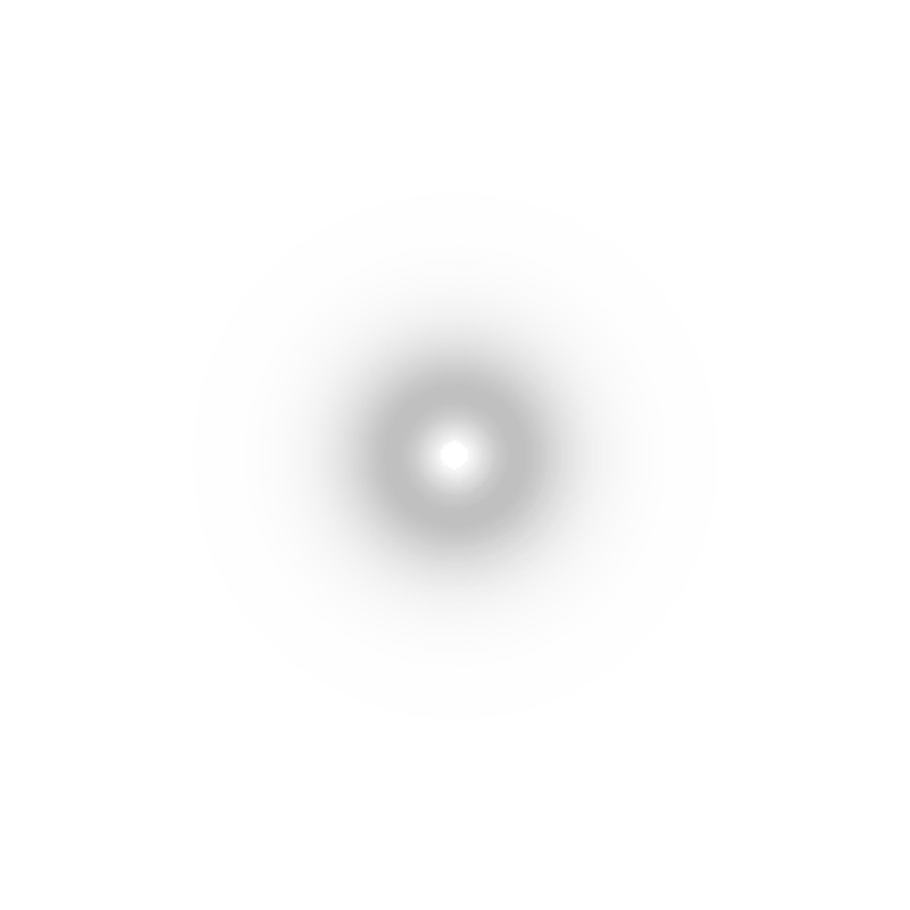}
\caption{The planar magnetic field, displayed in terms of $r/r_0$ for $r_0=1$ (left) and $2$ (right).}
\label{fig4}
\end{figure}

The profile of the magnetic field is similar to the one of the Chern-Simons-Higgs model, but in the current model the parameter $r_0$ can be used to control its intensity around $r_0$ itself. To show this more explicitly, in Fig.~\ref{fig4} we display the planar magnetic field for $r_0=1$ and $2$ to emphasize this behavior as one varies $r_0$. 

We may consider other possibilities for the generalized magnetic permeability, and now we take $P(\chi)$ in the form
\be
P_2(\chi)=\frac{1}{\chi^2}.
\ee
In this case, at the origin and asymptotically one has $\chi^2=1$, so $P_2(\chi)$ becomes unity and the vortex behaves as the Nielsen-Olesen one \cite{NO}. However, at $r=r_0$ the scalar field vanishes and makes $P_2(\chi)$ divergent, and this forces the magnetic field to vanish at $r=r_0$, introducing an internal structure to the vortex.

\begin{figure}[htb!]
\centering
\includegraphics[width=4.2cm]{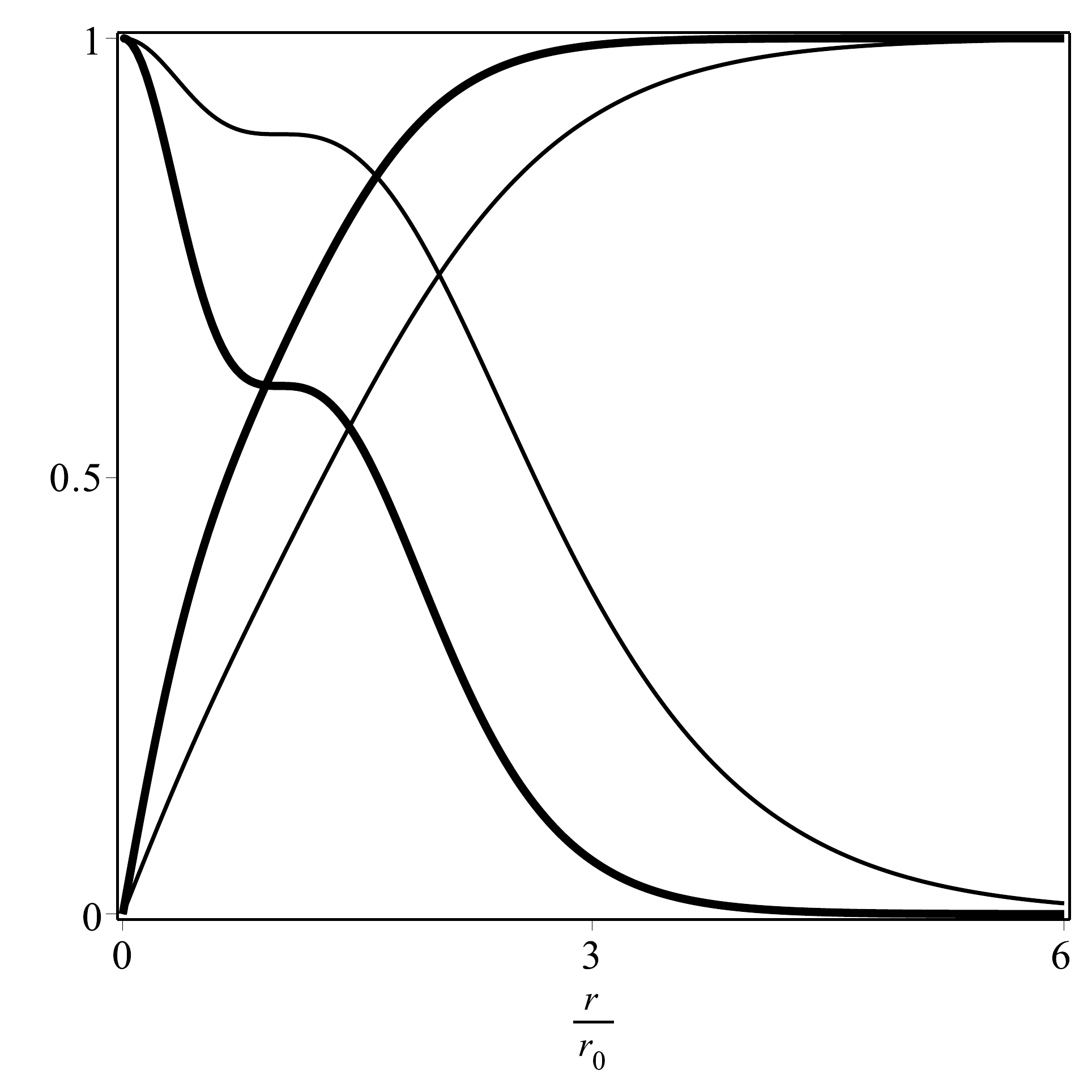}
\includegraphics[width=4.2cm]{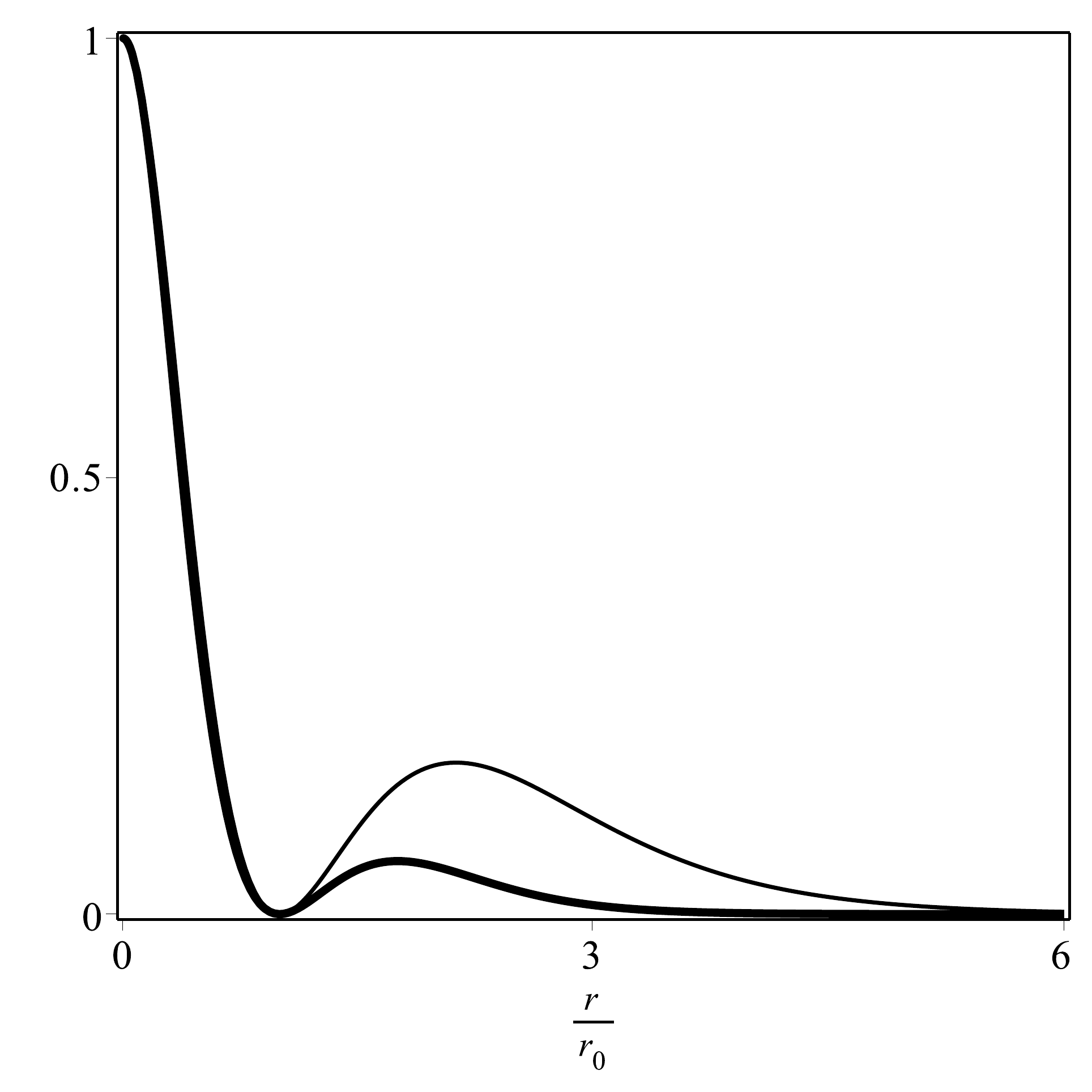}
\caption{The vortex solutions (left panel) $a(r)$ (descending lines) and $g(r)$ (ascending lines) and the magnetic field (right panel) in terms of $r/r_0$ for $r_0 = 1$ and $2$, with the thickness of the lines increasing with $r_0$.}
\label{fig5}
\end{figure}

In order to investigate the behavior of the vortex in this case, we need to solve equations \eqref{fog} and \eqref{foa}, the last one changing to
\be
-\frac{a^\prime}{r} = \frac{(r^2-r_0^2)^2\,(1-g^2)}{(r^2+r_0^2)^2}.
\ee
We use numerical procedures to investigate the system. In Fig.~\ref{fig5}, we display the solutions $a(r)$ and $g(r)$, and the magnetic field for some values of $r_0$, with the radial coordinate normalized to $r/r_0$. We see that the gauge field $a(r)$ has a different behavior around $r=r_0$, and this modifies the magnetic field accordingly, which now vanishes at $r=r_0$. 
\begin{figure}[htb!]
\centering
\includegraphics[width=4.2cm]{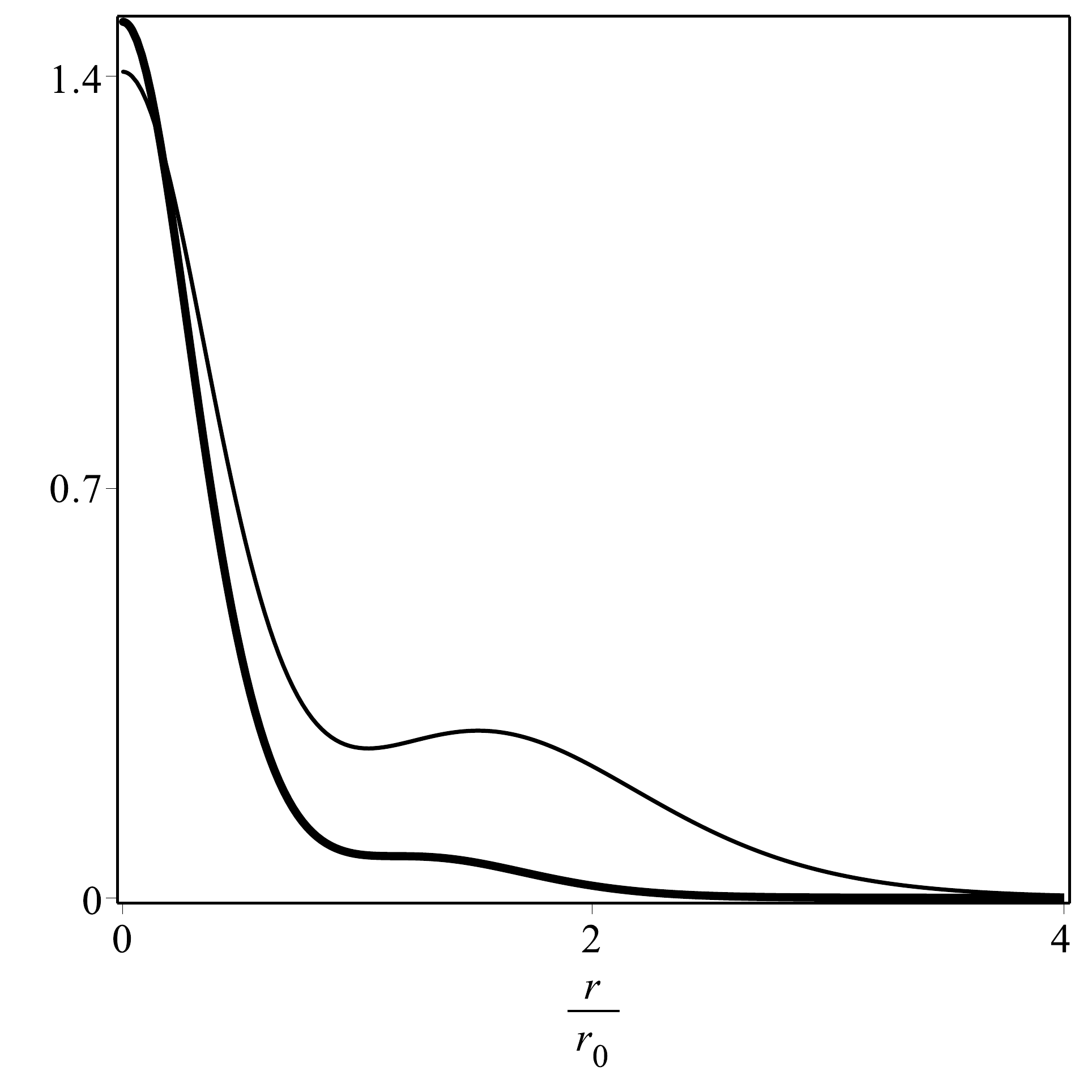}
\includegraphics[width=4.2cm]{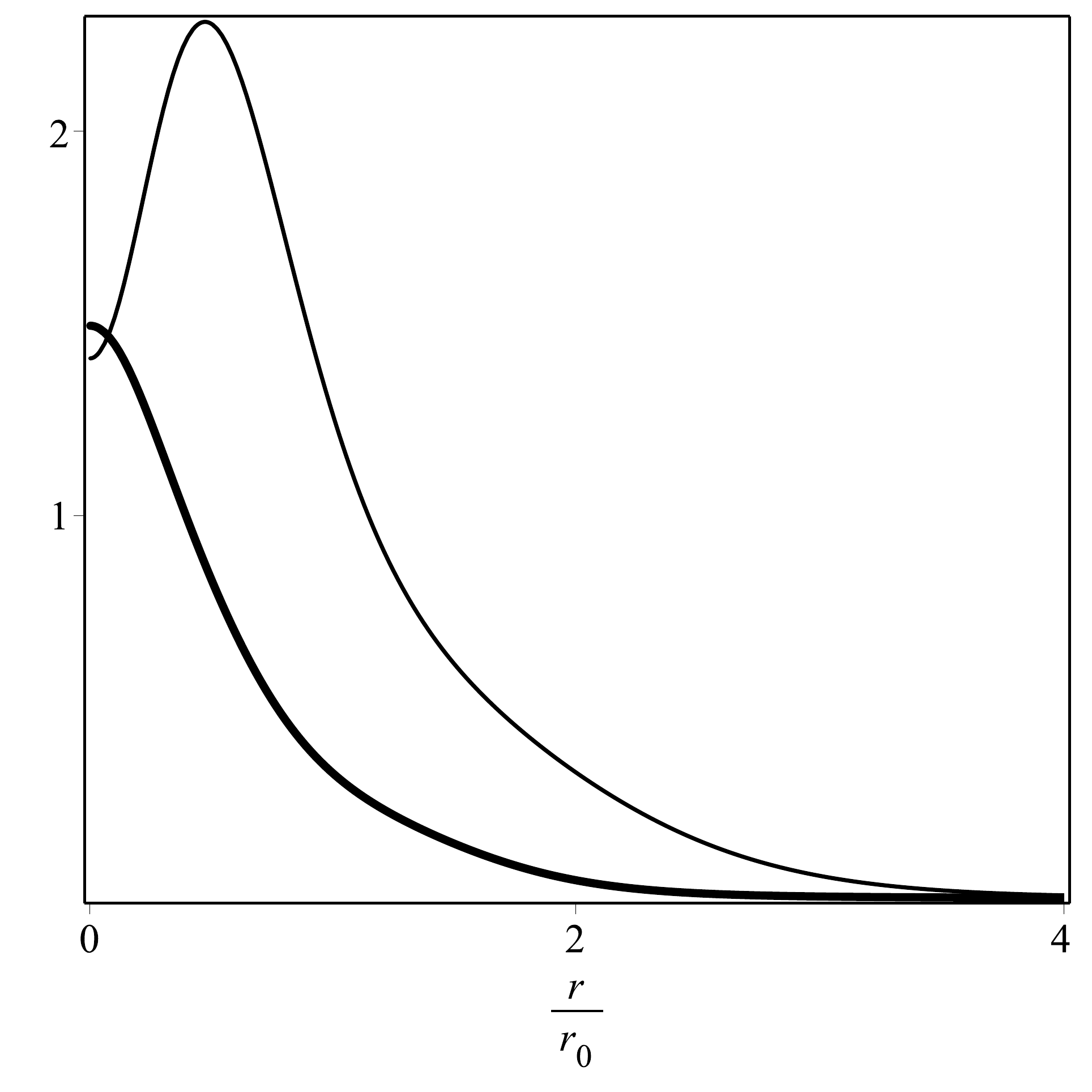}
\caption{The energy densities of the vortex (left panel) and the total energy density (right panel) for $r_0=1$ and $2$, with the thickness of the lines increases with $r_0$.}
\label{fig6}
\end{figure} 
\begin{figure}[t]
\centering
\includegraphics[width=4.2cm]{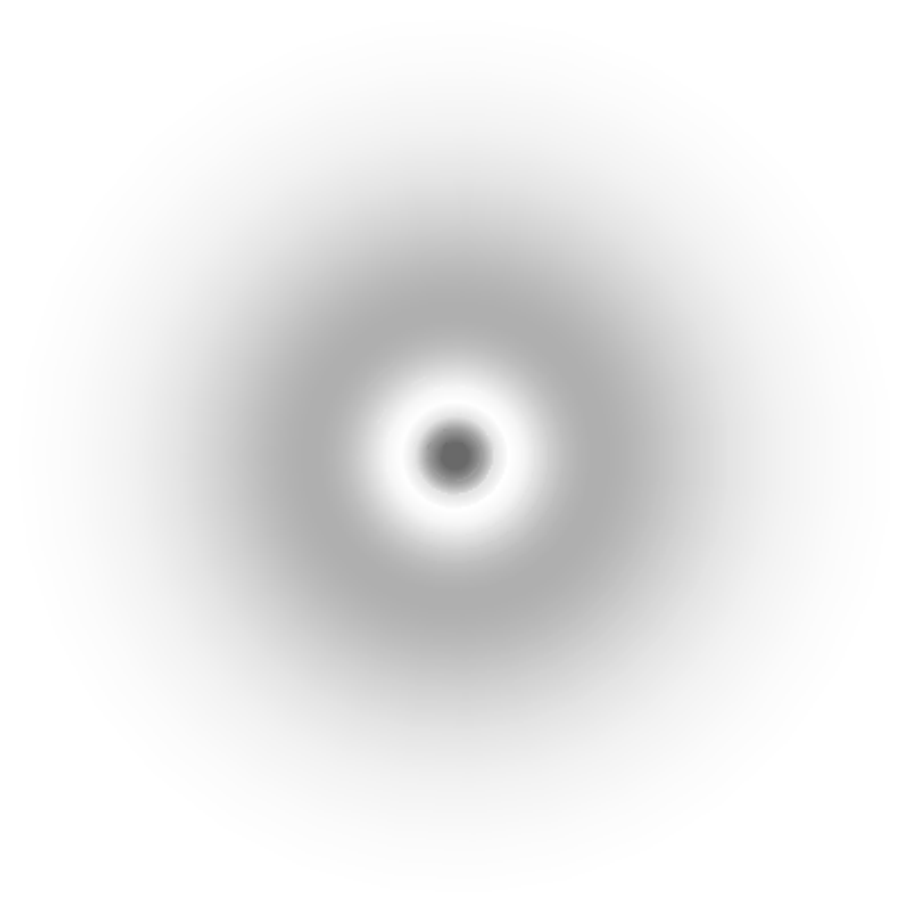}
\includegraphics[width=4.2cm]{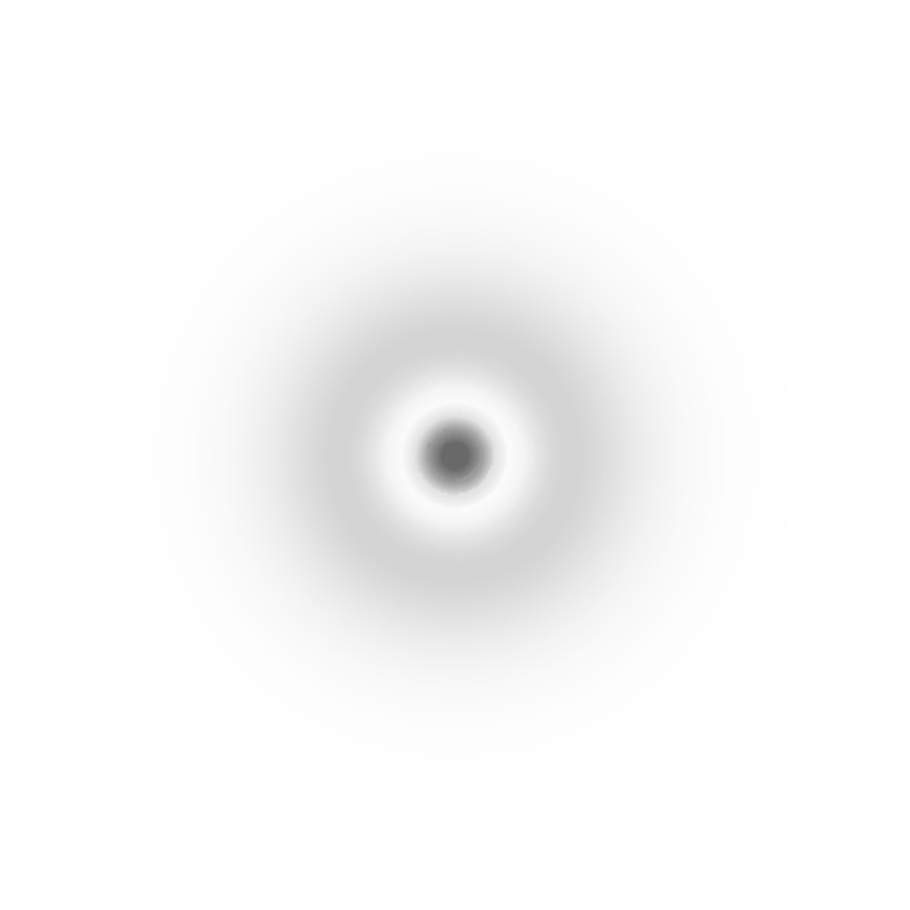}
\caption{The planar magnetic field, displayed in terms of $r/r_0$ for $r_0 =0.5$ (left) and $1$ (right).}
\label{fig7}
\end{figure} 

The energy density of the vortex can be calculated as before, and in Fig.~\ref{fig6} we depict it and the total energy density of the field configurations as functions of $r/r_0$ for some values of $r_0$. Also, in Fig.~\ref{fig7} we depict the planar magnetic field for $r_0=0.5$ and $1$, to better emphasize its novel behavior: it decreases from unity to zero at $r=r_0$, and then increases and decreases towards zero asymptotically. In the interval $[0,r_0]$ it remembers the solution of the standard Maxwell-Higgs model, and in the interval $[r_0,\infty)$ it behaves as in the Chern-Simons-Higgs model. This is the profile of a vortex with internal structure.

{\it Ending comments.} In this work we investigated a Maxwell-Higgs model in $(2,1)$ spacetime dimensions, with the addition of a neutral field that interacts with the gauge field via the inclusion of a generalized magnetic permeability. The neutral field also interacts with the charged scalar field via the Higgs-like potential. We have chosen the potential in a way that makes the energy of the field configurations to be minimized to its Bogomol'nyi bound, and this has led us to three first order differential equations that solve the equations of motion.

Interestingly, the first order equation of the neutral field decouples from the other two equations and can be solved independently. This makes the  neutral field the source field to generate the vortex configuration, and we have studied two distinct possibilities. In one case, the magnetic field of the vortex acquires the profile of the Chern-Simons-Higgs model. In the other case, the magnetic field seem to describe a vortex with internal structure.

The last possibility is new and may find applications in a diversity of contexts of current interest in nonlinear science. In the case of domain walls, for instance, it reminds us of the Bloch wall, which may be seem as an Ising wall with internal structure. The generalized model that we investigated in this work may be generated in metamaterials, and there it may find applications of current interest; see, e.g., Refs.~\cite{M1,M2,M3} and references therein. The novel vortex configuration may also appear in dipolar atomic Bose-Einstein condensates, when the magnetic dipole moments of the atoms effectively participate of the atomic interaction; see, e.g., Refs.~\cite{D1,D2,D3} and references therein. Moreover, the vortices with internal structure obey first order equations that minimize the energy, so they seem to be immersed in the bosonic portion of a larger, supersymmetric theory. Also, in the context of a larger theory, involving the $U(1)\times U(1)$ symmetry with visible and hidden sectors, the inclusion of generalized magnetic permeabilities opens new possibilities of study of current interest in high energy physics, as commented before in \cite{ana2}, for instance.

\acknowledgements{We would like to acknowledge the Brazilian agency CNPq for partial financial support. DB thanks support from grant 306614/2014-6, MAM thanks support from grant 140735/2015-1 and RM thanks support from grant 306826/2015-1.}

\end{document}